\documentclass [useAMS]{mn2e}
\usepackage{mathrsfs}
\usepackage {graphicx}
\begin{document}

\title[Magnetically Induced Disc-Corona]{A Model of Magnetically Induced Disc-Corona for
Black Hole Binaries }

\author[Z.-M. Gan, D.-X. Wang \& W.-H. Lei]
{Zhao-Ming Gan, Ding-Xiong Wang\thanks{Send offprint requests to: D.-X. Wang(dxwang@hust.edu.cn)} and Wei-Hua Lei\\
School of Physics, Huazhong University of Science and Technology,
Wuhan 430074, China }


\maketitle

\begin{abstract}
We propose a model of magnetic connection (MC) of a black hole with
its surrounding accretion disc based on large-scale magnetic field.
The MC gives {rise} to {transport} of energy and angular momentum
between the black hole and the disc, and the {closed} field lines
pipe the hot matter evaporated from the disc, and shape it in the
corona above the disc to form a magnetically induced disc-corona
system, in which the corona has the same configuration as the
large-scale magnetic field. We numerically solve the dynamic
equations in the context of the Kerr metric, in which the
large-scale magnetic field is determined by dynamo process and
equipartition between magnetic pressure and gas pressure. Thus we
can obtain a global solution rather than assuming the distribution
of large-scale magnetic field beforehand. The main MC effects lie in
three aspects. (1) The rotational energy of a fast-spinning black
hole can be extracted, enhancing the dissipation in the accretion
disc, (2) the {closed} field lines provide a natural channel for
corona matter escaping from disc and finally falling into black
hole, and (3) the scope of the corona can be bounded by the
conservation of magnetic flux. We simulate the high-energy spectra
of this system by using Monte-Carlo method, and find that the
relative hardness of the spectra decreases as accretion rate or
black hole spin $a_*$ increases. We fit the typical X-ray spectra of
three black-hole binaries (GRO J1655-40, XTE 1118+480 and GX 339-4)
in the low/hard or very high state.
\end{abstract}


\section{ INTRODUCTION}

It is well known that black hole binary transients present different
spectra in different stages during their outbursts, which are
usually referred to as low/hard state, high/soft state and
intermediate (or very high) state (McClintock \& Remillard 2006,
hereafter MR06). It is commonly realized that how to interpret the
different spectral profiles and the inducement of state transitions
may be the key to look into the real physical processes during the
outbursts.

Accretion disc is generally considered as the main energy source for
compact objects. Historically, people proposed various accretion
models to interpret different observed spectra and light curves,
such as standard thin disc (Shakura \& Sunyaev 1973, hereafter SS73;
Pringle 1981) and advection dominated accretion flow (ADAF, Narayan
\& Yi 1994). The main criterions to define different spectral states
are the relative weight of power-law component to thermal component
in X-ray spectra, and the photon index of power-law component
(MR06). However, we cannot expect to interpret the observed various
spectra based on a single accretion mode. For example, standard thin
disc is largely good for soft spectra of high/soft state, but it
cannot interpret the hackneyed power-law component in high energy
spectra. ADAF is congenitally propitious to hard spectra of low/hard
state, but it is invalid in fitting the observed soft excess and
also inverse radio spectra. Recently, it is generally considered
that different spectral components have different origins, i.e., the
thermal component of X-ray spectra comes from the black-body
radiation of a standard thin disc, while the power-law component
arises from corona or jet. And radio emission is usually considered
to be dominated by the synchronization radiation from jet.

Angular momentum {transport} is the essence of accretion theory.
Shakura \& Sunyaev (1973) proposed the seminal alpha viscosity law
to discuss complicated viscous process. Balbus \& Hawley (1991)
argued that magnetic rotation instability of tangled magnetic field
within accretion disc would be the physical origin of viscosity, and
it is generally regarded as another milestone of accretion theory.
On the other hand, the heating mechanism for corona is another open
question, in which magnetic field plays an essential role. The most
popular sketch is that the tangled small-scale magnetic field
amplified by dynamo process buoys up into corona and then
reconnects, heating corona by magnetic energy dissipation.
Meanwhile, the collimation mechanism of observed relativistic jets
is also under controversy. Recently, Miller et al. (2008) carefully
investigated the origin of disc wind of GRO J1655-40 during its
outburst in 2005.  Interestingly, he ruled out several popular
sketches for launching disc wind and concluded that the disc wind of
GRO J1655-40 must be driven by magnetic field. More and more
attention has been paid to the important role of magnetic fields in
spectral states of black hole binaries.

As is well known, the Blandford-Znajeck (BZ) process is an important
mechanism for jet production, in which energy is extracted
efficiently from a spinning black hole to power remote astrophysical
loads by invoking "open" large-scale magnetic field and frame
dragging effect. The BZ process can be regarded as the magnetic
Penrose process (Penrose 1969; Blandford \& Znajeck 1977; Livio,
Ogilvie \& Pringle 1999). As a variation of the BZ process, {closed}
field lines connecting a black hole with its surrounding accretion
disc lead to the transfer of energy and angular momentum between the
hole and the disc, and it is referred to as the Magnetic Connection
(MC) process (Li 2002; Wang et al. 2002). The direction of such
{transport} is determined by the angular velocity of the hole
relative to that of the disc. Energy can be extracted efficiently
from a spinning black hole with $a_* > 0.3594$ to the inner
accretion disc in the MC process, enhancing disc radiation and a
very steep emissivity index (Wilms et al. 2001; Wang et al. 2003).

{However, the importance of the MC remains controversial. Tomimatsu
\& Takahashi (2001) obtained a global magnetospheric structure with
a closed-loop and open field lines threading the inner and outer
parts of the disc, and the poloidal magnetic field is generated by a
toroidal electric current in a thin disc with an inner edge based on
the vacuum Maxwell equations in the Schwarzschild background. They
found that the MC between the black hole and the disc breaks down if
a uniform external field is strong enough. Hirose et al. (2004,
hereafter H04) presented a detailed analysis of the magnetic field
structure based on a set of three-dimensional general relativistic
magnetohydrodynamic (GRMHD) numerical simulations of accreting tori
in the Kerr metric with different black hole spins. Very recently,
Fragile \& Meier (2008, hereafter FM08) presented one of the first
physically-motivated two-dimensional GRMHD numerical simulations of
a radiatively-cooled black hole accretion disc, and they obtained a
magnetically-dominated accretion flow (MDAF) in the interior with an
outer advection-dominated accretion flow (ADAF). It turns out that
the importance of the MC depends on the radiative cooling, which is
not taken into account in the non-radiative simulations of H04. As
argued in FM08, a more ordered magnetic field could be formed due to
the radiative cooling, by which a dramatic increase in the dominance
of magnetic stresses is created. Considering the fact that a
standard thin disc with a very effective radiative cooling is
adopted in our model, we expect that the MC configuration of the
closed magnetic field lines can be formed.}

{Another controversial issue involves the stresses on the inner edge
of an accretion disc. Some authors (Krolik 1999; Gammie 1999; Agol
\& Krolik, 2000) argued that the torques exerted on the inner edge
of a disc might arise from the MC to the plunge region. Hawley and
Krolik (2002) showed that magnetic stresses between the plunge
region and the inner disk grow 10 times the value of stresses in the
disk body. Garofalo \& Reynolds (2005) showed the importance of such
inner disk torques and how they modify the dissipation profile of a
standard accretion disk even for Schwarzschild black holes. On the
other hand, Paczynski (2000) and Afshordi \& Paczynski (2003)
suggested that the zero-torque condition is likely to be a good
approximation for geometrically thin discs. This result was
confirmed by Shafee et al. (2008) who, using a global
height-integrated model, showed that modifications to the stress
profile are negligibly small for disk thicknesses $h$ less than
about a tenth of the local radius $r$. Since only a thin disc with
corona is involved in our model, we neglect the stresses on the
inner edge of the accretion disc for simplicity. }

Corona is a layer of hot tenuous plasma with temperature $ T_e >\sim
10^9K $ above accretion disc. The soft photons emitted from disc
surface are scattered up by relativistic electrons in corona,
resulting in a much harder spectrum. Corona is usually invoked to
interpret the power-law spectral component from tens of KeV to
hundreds of KeV observed in black hole binaries. The main
differences between various corona models lie in the heating
mechanism and the configuration of corona (i.e., temperature,
density and geometry). Stella \& Nosner (1984) argued that tangled
magnetic field amplified by dynamo process could float into corona
and reconnect there, thus corona is heated by the dissipated
magnetic energy. With the assumed corona geometry, Liu, Mineshige \&
Shibata (2002, hereafter LMS02) proposed a disc-corona system by
resolving equation of energy with the balance between thermal
conduction and corona evaporation. It is noted that coronae make
spectra harder, but cannot augment total luminosities. In addition,
Merloni \& Fabian (2002, hereafter MF02) pointed out that corona is
also a perfect launching site for outflow from accretion disc.\\

In this paper, we propose a magnetic disc-corona model, in which the
magnetic field configuration is determined self-consistently, and we
discuss the effects of magnetically induced corona on spectra of a
relativistic thin Keplerian disc. Based on a few assumptions, we
obtain a global solution of dynamic equations of accretion disc, by
which the interaction of corona, the origin of large-scale magnetic
field and the MC process are clarified. In this model, the MC
process not only transfers rotational energy from black hole to the
inner disc, but also provides a natural tube for evaporated hot
plasma to fall into the hole {as argued in FM08}. The tenuous hot
plasma is constrained and forced to move along the magnetic field
lines, giving {rise} to corona naturally according to the
configuration of the {closed} field lines. We assume that corona is
of the same configuration as the {closed} magnetic field for the MC
process, and calculate the boundary radius of corona based on
evolving the magnetic field configuration. It turns out that the
disc-corona system overwhelmingly dominates the high-energy
radiation of whole system.

Both the tangled small-scale magnetic field within accretion disc
and the ordered large-scale magnetic field related to the MC process
are involved in this model. It is assumed that the energy density of
small-scale magnetic field and the gas pressure within the disc are
of the same order based on energy equipartition. Magnetic rotational
instability (MRI) or turbulence of tangled small-scale magnetic
field is expected to work for viscous dissipation, while extraction
and {transport} of energy and angular momentum arise mainly from the
large-scale magnetic field. The small-scale magnetic field can be
amplified by dynamo process, and corona is heated by magnetic
buoyancy and reconnection. Large-scale magnetic field is generated
from the small-scale magnetic field, and the former is scaled by the
disc radius and its intensity is usually $1\sim3$ order weaker than
the latter. We present a numeric code to solve the coupled dynamic
equations of this magnetic disc-corona system self-consistently and
simulate its spectra by using Monte-Carlo method.

This paper is organized as follows. We outline the model in \S 2,
where the coupled disc-corona system and the MC process are
included. In \S 3, we numerically resolve the magnetic disc-corona
system and simulate the high-energy spectra from the inner disc by
using {Monte-Carlo} method. And four sampling results are presented
with detailed analysis. In \S 4, we fit the typical X-ray spectra of
three black-hole binaries during their outbursts. {\S 5} is the
discussion with a brief conclusion.

The Boyer-Lindquist coordinates and geometric units $G=c=1$ are used
throughout this paper.

\section{ MODEL DESCRIPTION}

\subsection{ Disc and Corona}
In standard thin disc model, the interior viscous stress $t_{r\phi}$
is usually assumed to be proportional to total pressure $P_{tot}$
(including gas pressure, radiation pressure and magnetic pressure),
namely the famous alpha viscosity law (SS73). Viscous stress is
alternatively assumed to be proportional to gas pressure $P_{gas}$,
or radiation pressure $P_{rad}$ or the geometrical mean of gas
pressure and total pressure $\sqrt{ P_{gas} \cdot P_{tot}}$ and so
on (Stella \& Nosner 1984; Wandel \& Liang 1991; Taam \& Lin 1984;
Wang et al. 2004, table 2 and references therein). Wang et al.
(2004) fitted the X-ray luminosities of 56 radio-quiet AGNs and
concluded that the observations prefer to the interior viscous
stress proportional to gas pressure. If interior viscous process is
dominated by tangled small-scale magnetic field, the viscous
pressure should be comparable to magnetic pressure (Balbus \& Hawley
1991; MF02), and we assume

\begin{equation}
\label{eq1} - t_{r\varphi } = \alpha P_{gas} \sim P_{mag} = B_D^2 /
8\pi .
\end{equation}

According to typical disc-corona scenario, part of the viscously
dissipated energy $Q$ is released as $Q^{+}_{d}$ in the disc,
emitting eventually as black-body radiation and supplying seed
photons for Comptonization of corona. The rest dissipated energy,
$Q^{+}_{cor}$, heats corona and maintains its relativistic
temperature via magnetic reconnection. The quantity $Q^{+}_{cor}$ is
proportional to magnetic energy density and local Aflven speed
(MF02; LMS02), and we have

\begin{equation}
\label{eq2} Q = Q_d^ + + Q_{cor}^ + \equiv Q_{cor}^ + / f_{cor} ,
\end{equation}

\begin{equation}
\label{eq3} Q_d^ + = \sigma T_{eff}^4 , \quad Q_{cor}^ + =
\frac{B_D^2 }{4\pi }V_A = \frac{B_D^3 }{4\pi \sqrt {4\pi \rho } },
\end{equation}

\noindent where $B_D$, $V_A$ and $\rho$ are the intensity of
interior tangled small-scale magnetic field, local Alfven speed and
mass density, respectively. $T_{eff}$ is the effective temperature
of accretion disc and $\sigma$ is the Stefan-Boltzman constant.

In the context of relativistic hot corona, inverse Compton
scattering is very effective, by which corona cooling is dominated.
According to energy balance given by LMS02, $Q^{+}_{cor} = Q^{-}_
{comp}$, we have

\begin{equation}
\label{eq4} \frac{B_D^3 }{4\pi \sqrt {4\pi \rho } } =
\frac{4kT_{cor} }{m_e }\tau _{cor} \cdot U_{rad} ,
\end{equation}

\noindent where $U_{rad} = a T^4_{eff} = 4 Q^{+}_{d}$ is the radiant
energy density at vicinity of disc surface. $T_{cor}$ and
$\tau_{cor}$ are the temperature and the optical depth of corona.
The quantities $k$, $m_e$ and $a$ are the Boltzman constant,
electron mass and radiation constant, respectively. Observations
reveal that optical depths of corona for AGNs and for black hole
binaries in low/hard state (LHS) lie in a very narrow range around
$\tau_{cor} \sim 1$ (Gierlinski et al. 1997; Zdziarski 1999), and we
adopt $\tau _{cor}=1$ for facility, i.e.,

\begin{equation}
\label{eq5} \tau _{cor} = n_{cor} \sigma _T l\sim 1,
\end{equation}

\noindent where $n_{cor}$ and $l$ are the number density of electron
and the height of corona, respectively.

\subsection{ The MC Process}

Being different from standard thin disc, energy dissipation $Q$ in
this model arises from two sources: one is the gravitationally bound
energy of accreting matter, and the other is the rotational energy
extracted from the black hole via the MC process (Wang et al. 2002).

A fast-spinning black hole exerts a significant torque on its
surrounding disc, if the two are coupled by the closed magnetic
field. Energy and angular momentum are transferred between the two
simultaneously. The torque exerted by the black hole onto an
infinitesimal annular region of the disc is given by

\begin{equation}
\label{eq6} dT_{MC} = 2\left( {\frac{d\psi }{2\pi }}
\right)^2\frac{(\Omega _H - \Omega _D )}{dZ_H } \equiv 4\pi rdr
\cdot H_{MC} ,
\end{equation}

\noindent where $H_{MC} \equiv (1/4\pi r) \cdot dT_{MC}/dr$ is the
flux of angular momentum transferred from the black hole to the
disc. The quantities $\Omega_H$ and $\Omega_D$ are the angular
velocity of the black hole and that of the disc, respectively. The
quantity $\psi=\psi(r,\theta)$ is the magnetic flux through a
surface bounded by a {curve} with $r=constant$ and
$\theta=constant$, and we have

\begin{equation}
\label{eq7} d\psi=B_p \cdot 2\pi r\sqrt {\frac{\mathcal
{A}}{\mathcal {D}}} \cdot dr \mid_{\theta=\pi/2}=-B_H \cdot 2\pi r_H
\cdot 2M\sin \theta d\theta \mid_{r=r_H},
\end{equation}

\noindent $r_H$ is the radius of black-hole horizon and the
resistance of the horizon corresponding to the magnetic flux is
written as (Wang et al. 2002)

\begin{equation}
\label{eq8} dZ_H = \left( {\frac{R_H }{2\pi }} \right)\frac{r_H^2 +
M^2a_\ast ^2 \cos ^2\theta }{(r_H^2 + M^2a_\ast ^2 )\sin \theta
}\left( {\frac{d\theta }{dr}} \right) \cdot dr
\end{equation}

\noindent $R_H$ is the surface resistivity of black-hole horizon.
$M$ and $a_*$ are the mass and the dimensionless spin of black hole,
respectively.

In equation (\ref{eq7}), $B_p$ is the poloidal component of the
large-scale magnetic field anchored on the disc. Equation
(\ref{eq7}) is derived from conservation of magnetic flux, by which
the mapping relation $r(\theta)$ between black-hole horizon and the
inner region of the disc can use determined. Assuming that the inner
edge of the disc connects the horizon at $\theta = \pi/2$, we obtain
the outer boundary of the MC region, $r(\theta = 0)$, by integrating
equation (\ref{eq7}) from $\theta = \pi/2$ to 0.

Still, we have to estimate magnetic field strength on the horizon
required by equations (\ref{eq6}) and (\ref{eq7}). Assuming a
uniform magnetic field on the horizon and adopting the equilibrium
between magnetic pressure and the ram pressure given by Moderski,
Sikora \& Lasota (1997), we have

\begin{equation}
\label{eq9} B_H = \sqrt {{2\dot {M}_D } \mathord{\left/ {\vphantom
{{2\dot {M}_D } {r_H^2 }}} \right. \kern-\nulldelimiterspace} {r_H^2
}} ,
\end{equation}

\noindent where $\dot{M}_D$ is accretion rate.

The relativistic conservation equations of energy and angular
momentum for a disc with magnetic {connection} can be written as (Li
2002)

\begin{equation}
\label{eq10} \frac{d}{dr}(\dot {M}_D L^\dag - g) = 4\pi r(QL^\dag -
H_{MC} ),
\end{equation}

\begin{equation}
\label{eq11} \frac{d}{dr}(\dot {M}_D E^\dag - g \cdot \Omega _D ) =
4\pi r(QE^\dag - H_{MC} \Omega _D ).
\end{equation}

The quantities $Q$ and $g$ are the dissipated energy per unit disc
surface and interior viscous torque of the disc, respectively. And
$E^\dag$, $L^\dag$ and $\Omega_D$ are respectively the specific
energy, specific angular momentum and angular velocity of the test
particles in the equatorial plane of a Kerr black hole (Bardeen,
Press \& Teukolsky 1972).

\begin{equation}
\label{eq12}
\begin{array}{l}
\Omega_D = {\sqrt{{M}/{r^3}}}\ /{(1 + a_\ast /
\tilde {r}^\frac{3}{2})},\\

E^\dag =\frac{(1-2/\tilde{r}+a_{\ast}/{\tilde{r}^\frac{3}{2}})}
{\sqrt{1-3/\tilde{r}+2a_\ast/\tilde {r}^\frac{3}{2}}},\

L^\dag=\sqrt{Mr} \cdot
\frac{(1-{2a_\ast^2}/{\tilde{r}^\frac{3}{2}}+{a_\ast^2}/{\tilde{r}^2})}
{\sqrt{1-3/\tilde{r}+2a_\ast/\tilde {r}^\frac{3}{2}}},
\end{array}
\end{equation}

\noindent where $\tilde{r} \equiv r/M$. The quantities $\mathcal
{A}$ and $\mathcal {D}$ in equation (\ref{eq7}) are relativistic
correction factors in Kerr metric, and they read

\begin{equation}
\label{eq13} \ \  \mathcal{A}=1+a_\ast ^2 / \tilde{r}^2+2a_\ast^2 /
\tilde{r}^3,\quad  \mathcal{D}=1-2 / \tilde{r}+a_\ast^2 /
\tilde{r}^2
\end{equation}

\ \newline We can derive the energy dissipation $Q$ and interior
viscous torque $g$ by resolving equations (\ref{eq10}) and
(\ref{eq11}).

\begin{equation}
\label{eq14}
\begin{array}{l}Q(r) \equiv Q_{DA} + Q_{MC} = Q_{DA} + \\
\quad \quad \quad \frac{1}{r}\frac{ - d\Omega _D / dr}{(E^\dag -
\Omega _D L^\dag )^2} \cdot \int_{r_{ms} }^r {(E^\dag - \Omega _D
L^\dag )} H_{MC} rdr
\end{array}
\end{equation}

\begin{equation}
\label{eq15} \ \  Q_{DA} = \frac{1}{4\pi r}\frac{ - d\Omega _D /
dr}{(E^\dag - \Omega _D L^\dag )^2}\int_{r_{ms} }^r {(E^\dag -
\Omega _D L^\dag )} \dot {M}_D \cdot dL^\dag
\end{equation}

\begin{equation}
\label{eq16} \ \ g(r) = \frac{E^\dag - \Omega _D L^\dag }{ - d\Omega
_D / dr}4\pi r \cdot Q(r)
\end{equation}

\noindent $r_{ms}$ is the radius of the innermost stable circular
orbit, which is assumed to be the inner edge of accretion disc.

In this model we intend to resolve the dynamic equations combining
the MC effects, which might change the dynamic property of accretion
disc significantly. To resolve equations (\ref{eq14})-(\ref{eq16}),
we adopt the relation between the large-scale magnetic field above
the disc and tangled small-scale magnetic field in the disc as
follows (Livio et al. 1999),

\begin{equation}
\label{eq17} B_p \sim \left( {\frac{h}{r}} \right) \cdot B_D ,
\end{equation}

\noindent where $h$ is the half height of the disc. The intensity
and distribution of $B_D$ and $B_p$ can be derived by resolving the
dynamic equations.

\section{ NUMERICAL RESULTS}

For facility, we consider a slab corona with $l \sim 10 r_{ms}$.
Inspecting equations (\ref{eq6})-(\ref{eq16}) we find that magnetic
connection at radius $r$ could affect the disc region beyond $r$. We
design an iterative algorithm and integrate equations
(\ref{eq10})-(\ref{eq16}) from the innermost stable circular orbit
(ISCO) to the outer boundary of the MC region, and obtain
self-consistently global solutions of the disc-corona system, in
which the full solutions of corona and the MC effects are included.
As argued above, energy can be extracted from a fast-spinning black
hole via the MC process, resulting in the enhancement of dissipation
in the disc. Subsequently, the disc height, pressure and magnetic
field increase. It is noted that the MC effects scaled by the square
of magnetic field become very strong for the amplified magnetic
field. Since corona heating via magnetic reconnection is scaled by
the cube of magnetic field [cf. equation(\ref{eq3})], we expect that
a stable saturated magnetic configuration can be maintained in our
model.

The iterative algorithm consists of the following steps: (1)
resolving numerically the disc-corona system without the MC process;
(2) deriving the large-scale magnetic field and calculating the
energy extraction by the MC process; (3) combining the energy
dissipation of the disc with the MC process, and resolving the
disc-corona system; (4) repeating steps (2) and (3) until the
magnetic field is saturated; (5) integrating outward to a larger
radius ($r \sim r+dr$); (6) repeating (1)-(5) until the outer
boundary of MC process ( $\theta (r)=0$ ) is reached.

Here we present four typical solutions as shown in Figure 1-2, 3-4,
5-6 and 7-8, where $T_d$ is the temperature of accretion disc on the
equatorial plane. The concerned parameters are listed in Table 1, in
which the black-hole mass $m_{bh}$ ($\equiv M/M_\odot$), distance
$D$, inclination angle $i$, corona height $l$, viscous coefficient
$\alpha$, black-hole spin $a_*$ and accretion rate $\dot{m}$ (in
unit of the Eddington rate) are included. The first three parameters
are constrained by observations. The viscous coefficient is taken as
a typical value, $\alpha = 0.3$. In fact the corona height might
vary with the disc radius, and its value could influence the density
of corona. Fortunately, emerged spectrum from corona is mainly
determined by the temperature and optical depth. Since we adopt a
constant optical depth for corona, our results are not sensitive to
$l$, and we fix $l = 10 r_{ms}$ in calculations. So there are
actually only two free parameters in our model, accretion rate
$\dot{m}$ and black-hole spin $a_*$. The former represents the "fuel
supply", and the latter determines the radiant efficiency
(luminosity per unit mass). Both parameters are very important for
the emerged spectrum. Actually, we can demonstrate that they are the
key elements of the spectral profile.\\

\begin{table*}
\begin{center}
\caption{Concerned Parameters in the sampling results and
simulations}

\begin{tabular}{cccccccccc}
\hline \hline

{Source} & \multicolumn{3}{c}{Observation}
&& \multicolumn{4}{c}{Model} & {{Results}}\\

\cline{2-4}  \cline{6-9}

&{$m_{bh}$}& {$D$ /kpc}& {$i$ /degree}&& {$l$ /$r_{ms}$}&
{$\alpha$}& $a*$& {$\dot{m}$}& {(Figure)} \\ \hline

Sample 1& 10.0& 3.0& 0.0&& 10.0& 0.3& 0.900& 0.300&1, 2 \\ 
Sample 2& 10.0& 3.0& 0.0&& 10.0& 0.3& 0.200& 0.030&3, 4 \\ 
Sample 3& 10.0& 3.0& 0.0&& 10.0& 0.3& 0.200& 0.300&5, 6 \\ 
Sample 4& 10.0& 3.0& 0.0&& 10.0& 0.3& 0.900& 0.030&7, 8 \\ 

GRO J1655-40 (LHS)& 7.0$^{ a}$& 3.2$^{ b}$& 70.0$^{ c}$&& 10.0& 0.3&
0.200& 0.030& 9 \\ 

GX 339-4 (VHS)& 5.8$^{ d}$& 7.0$^{ e}$&12.0$^{ f}$&& 10.0& 0.3&
0.600& 0.250&10 \\ 

XTE J1118+480 (LHS)& 8.0$^{ g}$& 1.8$^{ h}$& 70.0$^{ i}$&& 10.0&
0.3& 0.400& 0.005& 11 \\ \hline

\end{tabular}
\end{center}
\label{tab1}

\textbf{References}: {(a) Orosz, J. A., Bailyn, C. D., 1997, ApJ,
477, 876; (b) Tingay, S. J., et al., 1995, Nature, 374, 141; (c)
Bailyn, C. D., Orosz, J. A., et al., 1995, Nature, 378, 157; (d)
Hynes, R. I., Steeghs, D., et al., 2003, ApJ, 583, L95; (e)
Zdziarski A. A., Poutanen J., et al., 1998, MNRAS, 301, 435; (f)
Miller J. M., Fabian A. C., et al., 2004, ApJ, 606, L131; (g, h)
McClintock J. E., Garcia M. R., et al., 2001, ApJ, 551, 147; (i)
Wagner R. Mark, Foltz C. B., et al., 2001, ApJ, 556, 42. }
\end{table*}









\textbf{The MC effects on accretion disc}. The MC process utilizes
the differential rotation between a black hole and its accretion
disc to transfer energy and angular momentum. As argued by Wang et
al. (2002), the MC effects become much stronger for greater black
hole spin. In this paper, we find that the MC effects on accretion
disc also depend on accretion rate. The distribution of {closed}
field lines and corona is very concentrated to the innermost region
of accretion disc ($<\sim 20 M$) for great accretion rate ($>\sim
0.2 \dot{M}_{Edd}$). Thus the energy pumped into the disc via the MC
process is comparable to the released gravitational energy and the
radiation efficiency of the system is enhanced significantly. As
shown in Figure 1a and 1e, both temperature and dissipation rate of
the disc are augmented obviously due to the MC effects. It is
interesting to note that the MC process has little influence on gas
pressure, but it strengthens radiation pressure significantly. As a
result, the inner region of disc becomes thick enough, being apt to
form large-scale magnetic field. On the other hand, the stronger
large-scale magnetic field is, the more efficient is MC process. So
the energy dissipation and the MC process promote each other in our
model. When accretion rate is small ($ <\sim 0.05 \dot{M}_{Edd}$),
the distribution of {closed} field lines and corona can extend to
$>\sim 100 M$. In this case, the magnetic field is too weak to give
{rise} to {obvious} effects on the disc dynamics, and the {closed}
field lines mainly provide a channel for corona matter escaping from
the disc and finally falling into the hole.

Recalling our previous researches, we conclude that {transport} of
energy and angular momentum via the MC process is generally
concentrated in a very narrow region ($\sim 10 M$) near the inner
disc boundary. However, it can change the total luminosity of the
whole
system significantly.\\

\textbf{Emerged spectra of the disc-corona system}. We find that
spectral profiles change obviously with both accretion rate and
black-hole spin. Generally, thermal radiation mainly comes from the
disc, and power-law component comes from corona, implying that the
higher corona temperature has the harder spectrum.

(1) The fraction of thermal component increases as accretion rate
increases. When accretion rate increases, the magnetic field and
thermal production are enhanced, resulting in much more soft photons
escaping from the surface of disc. Thus the corona temperature is
suppressed effectively by the cooling of Comptonization.

(2) The fraction of power-law component decreases as black hole spin
increases. The gravitational potential well becomes very deep as the
black hole spins fast. On the other hand, the stronger frame
dragging effects give {rise} to more energy transferred from a
spinning hole to the disc in the MC process. Both effects can
strengthen the radiation from the inner disc as well as suppress the
corona temperature.

Generally speaking, the total luminosity of the disc-corona system
is positively correlated to accretion rate and black-hole spin. We
find that the spectra are always dominated by thermal component when
accretion rate or black-hole spin is large (e.g., $\dot{M}_D > 0.3
\dot{M}_{edd}$, $ a_* > 0.9 $), and these results are in agreement
with those obtained by Esin, McClintock \& Narayan (1997).

\section{ COMPARISON TO OBSERVATIONS}

We try to fit the typical X-ray spectra of 3 black-hole binaries in
the low/hard or very high state, i.e., GRO J1655-40, XTE 1118+480
and GX 339-4. The data are taken from literatures (without error
bars): GRO J1655-40 (Brocksopp et al. 2006), XTE 1118+480 (Yuan et
al. 2005 and references therein), GX 339-4 (Miller et al. 2004).




As shown in Figures {9 and 11}, the hard X-ray spectra with low
luminosity (GRO J1655-40 and XTE 1118+480 in low/hard state) are
well fitted. In these simulations, the accretion rates are small,
gravitational binding energy is mildly released along the inner
region of accretion disc and most of the energy (larger than 80\%)
is dissipated in corona via magnetic reconnection (cf. Figure 3f).
And it turns out that the spectra are hard but dim. {Inspecting
Figure 10, we find that the soft X-ray spectrum with very high
luminosity of GX 339-4 in very high state can also be well fitted.
In this simulation, the accretion rate and black-hole spin are
large. Gravitational binding energy is released efficiently and
efficacy of MC process also becomes significant. Energy dissipation
is concentrated at the inner most region of the disc. Meanwhile,
there is only a small fraction of dissipated energy distributed to
corona (cf. Figure 1f). And it turns out that the spectrum is soft
but bright.}

However, there are excesses in soft X-ray band in {these
simulations}, when we try to fit the total X-ray spectra. Since it
is an embryo model, we cannot expect our model to fit the full
features of the spectra.

\section{DISCUSSION}

In this paper, we propose a magnetic disc-corona model, in which the
configuration of large-scale magnetic field and corona can be
self-consistently determined with several minimal assumptions. We
combine accretion disc, corona, origin of large-scale magnetic field
and the MC process into this model to investigate the interaction
among them. The global solutions are obtained numerically. The size
of corona can be self-consistently determined by resolving the
magnetic configuration of the MC process. It is found that the MC
process could change the properties of accretion disc significantly
for a large accretion rate. In addition, we simulate the emerged
spectra from the inner region of accretion disc by using Monte-Carlo
method.

Observations reveal that large-scale magnetic fields exist in
compact objects. However, the origin of large-scale magnetic field
is still under controversy. The popular treatment of large-scale
magnetic field is invoked magnetohydrodynamical (MHD) simulations.
It is particularly noted that Uzdensky (2004, 2005) investigated the
MC between a black hole and its surrounding disc, and he solved the
Grad-Shafranov equations to determine the configuration of
large-scale magnetic field in a force-free magnetosphere. Instead,
we use the phenomenal description of dynamo mechanism to avoid
complex calculations, and treat the origin of large-scale magnetic
field directly by invoking the dynamics of accretion disc. In our
model, the large-scale magnetic field, corona and accretion disc are
connected very closely, which cannot be separated from each other.

It turns out that in this model the emerged spectra change
significantly with accretion rate, naturally giving rise to hard
spectra with low luminosity and soft spectra with high luminosity.
This feature might be related to the state transitions of black-hole
binaries. As argued by Esin {et al.}(1997), the state transition
could be induced by the variation of accretion rate. As a matter of
fact, the key for state transitions is the accretion rate of the
inner disc region rather than the rate of mass transferred from a
donator star. The variation of the accretion rate of the inner disc
region mainly has two possibilities: (1) mass rate from donator star
changes; (2) a large fraction of accreting matter is thrown off the
disc, e.g. by the Blandford-Payne (BP) process, in the form of jet
or disc wind (Blandford \& Payne 1982). It seems reasonable that
most of accreting matter escapes from accretion disc before it drops
into black hole in the low/hard state of black-hole binaries.

The BP process is a very important mechanism for disc wind and jet
production, which is also related directly to large-scale magnetic
field. In the BP process, accreting matter could be thrown
centrifugally off accretion disc and accelerated along the field
lines. The outflow can be self-collimated to jet due to magnetic
pinch effect. However, it would become disc wind for weak magnetic
field. The BP process can change the accretion rate of the inner
disc region likely and thus affect the high energy emission of the
system. It might play a very important role in state transitions of
black-hole binaries. The outflow is also expected to produce some
radiation and absorptive effects. {Although there is little evidence
of BP winds in GRMHD (see e.g. H04), the observed rich absorptive
lines of GRO J1655-40 are likely related to winds driven by the BP
process (Miller et al. 2008).} The presence of BP process in our
model would help to fit the observations
and we leave it to the future work.  \\

\noindent {\bf Acknowledgements:}\quad This work is supported by the
National Natural Science Foundation of China under grant 10873005
and National Basic Research Program of China under grant
2009CB824800. {We are very grateful to the anonymous referee
for his (her) helpful comments on the manuscript.}\\

\newpage

\begin{figure*}
\label{fig1} \vspace{0.5cm}
\begin{center}

\includegraphics[height=10cm]{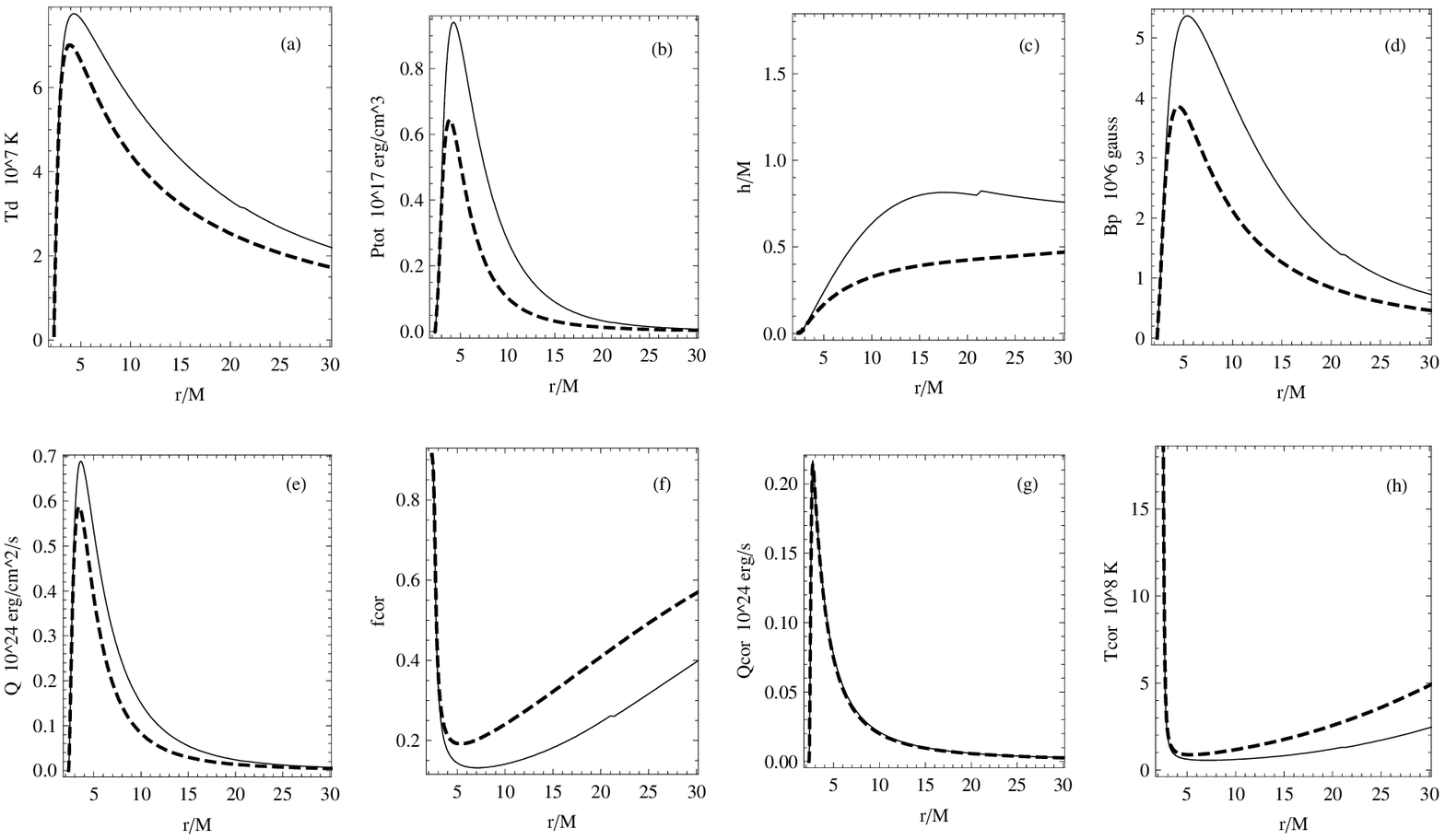}

\end{center}
\caption{A global solution of the magnetic disc-corona system in the
cases of large accretion rate and large black-hole spin (Sample 1).
Quantities are plotted in dashed and solid lines for disc + corona
and disc + corona + MC, respectively.}
\end{figure*}

\begin{figure*}
\label{fig1} \vspace{0.5cm}
\begin{center}

\includegraphics[height=8cm]{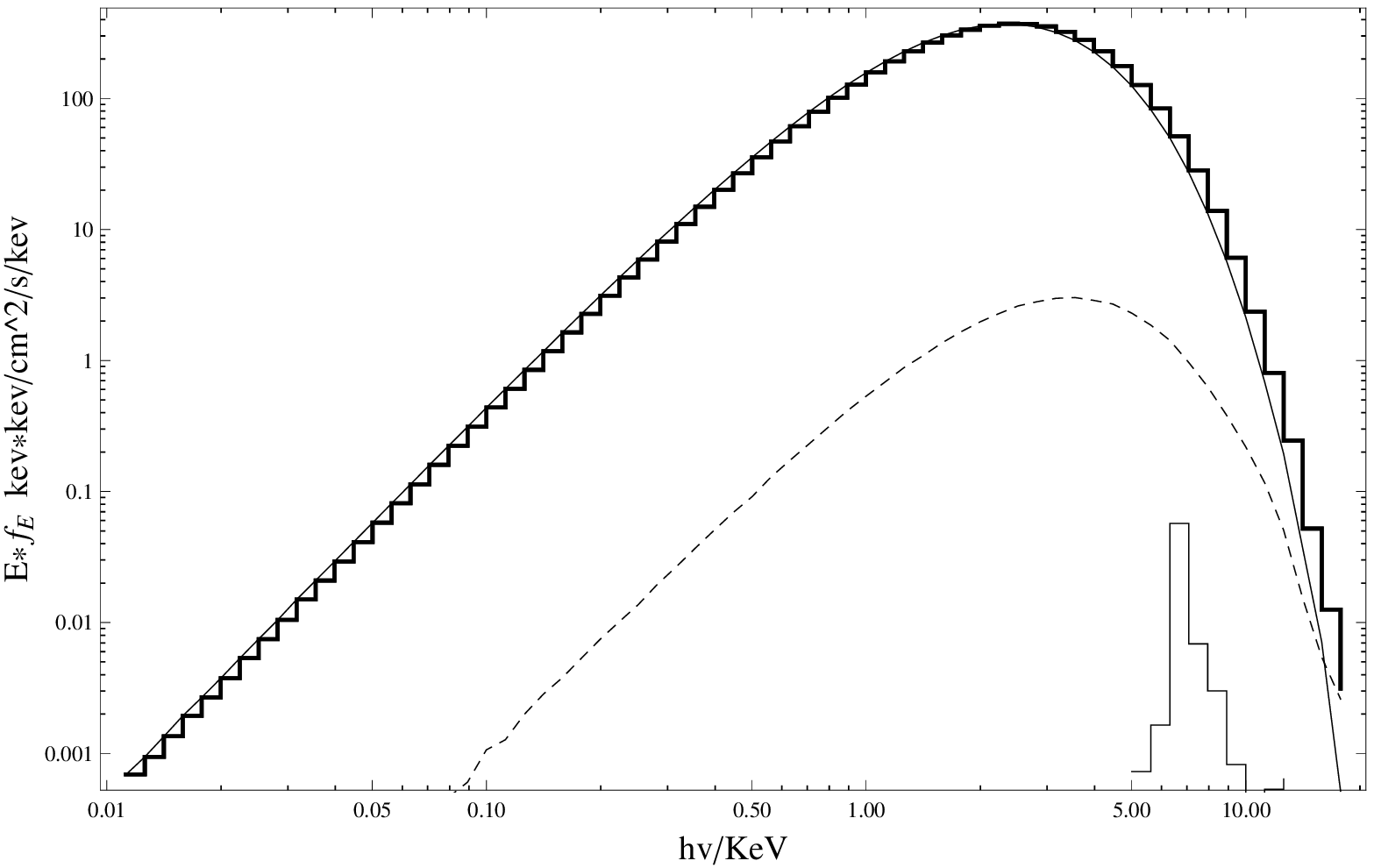}

\end{center}
\caption{Emerged spectrum from the disc-corona system in the case of
large accretion rate and large black-hole spin (Sample 1). The total
emissive spectrum and its thermal, comptonized and reflective
components are plotted in thick-zigzag, solid, dashed and
thin-zigzag lines, respectively.}
\end{figure*}

\begin{figure*}
\label{fig1} \vspace{0.5cm}
\begin{center}

\includegraphics[height=10cm]{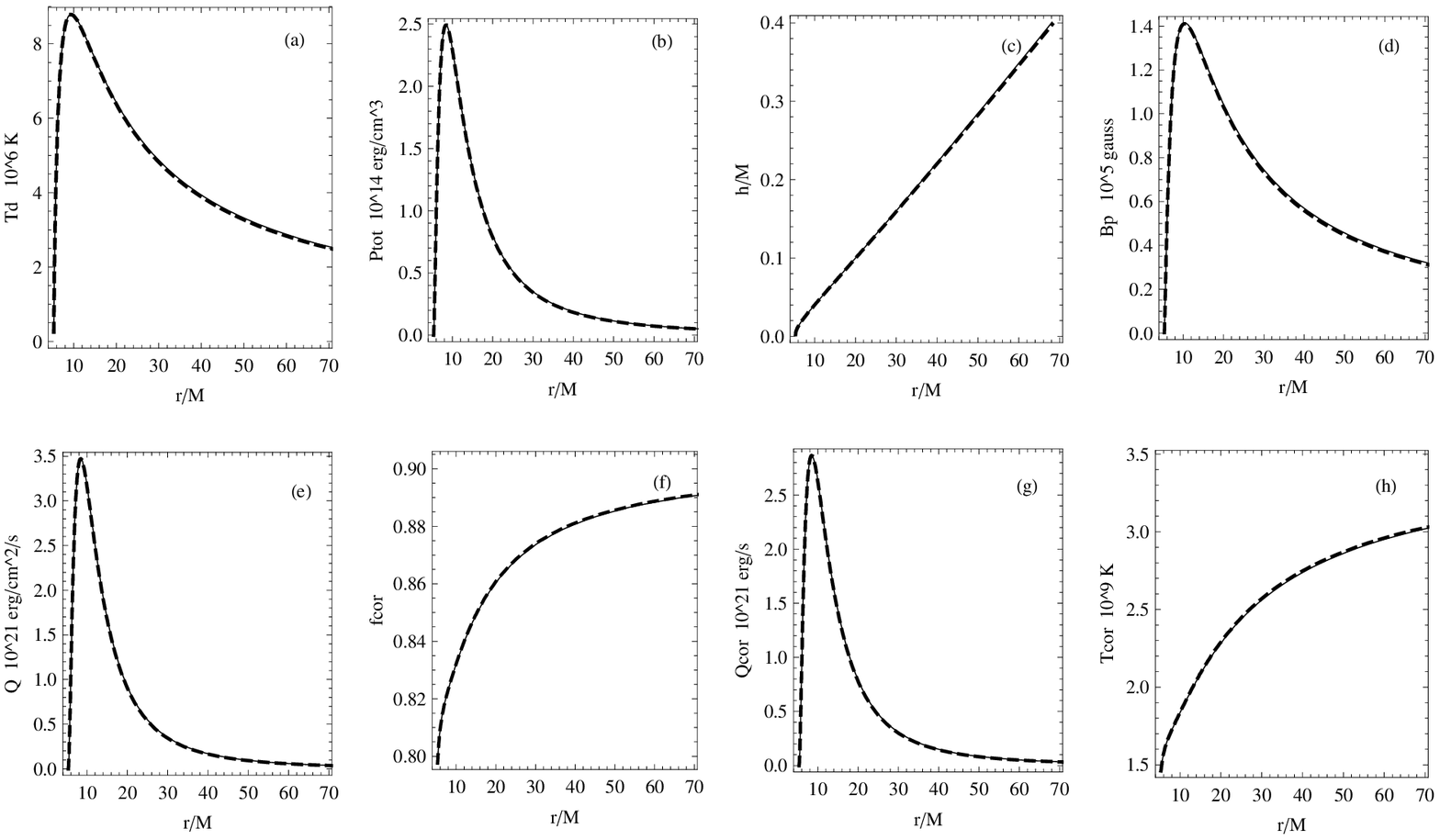}

\end{center}
\caption{A global solution of the magnetic disc-corona system in the
cases of small accretion rate and small black-hole spin (Sample 2).
Quantities are plotted in dashed and solid lines for disc + corona
and disc + corona + MC, respectively. (The dashed and solid lines
almost coincide completely.)}
\end{figure*}

\begin{figure*}
\label{fig1} \vspace{0.5cm}
\begin{center}

\includegraphics[height=8cm]{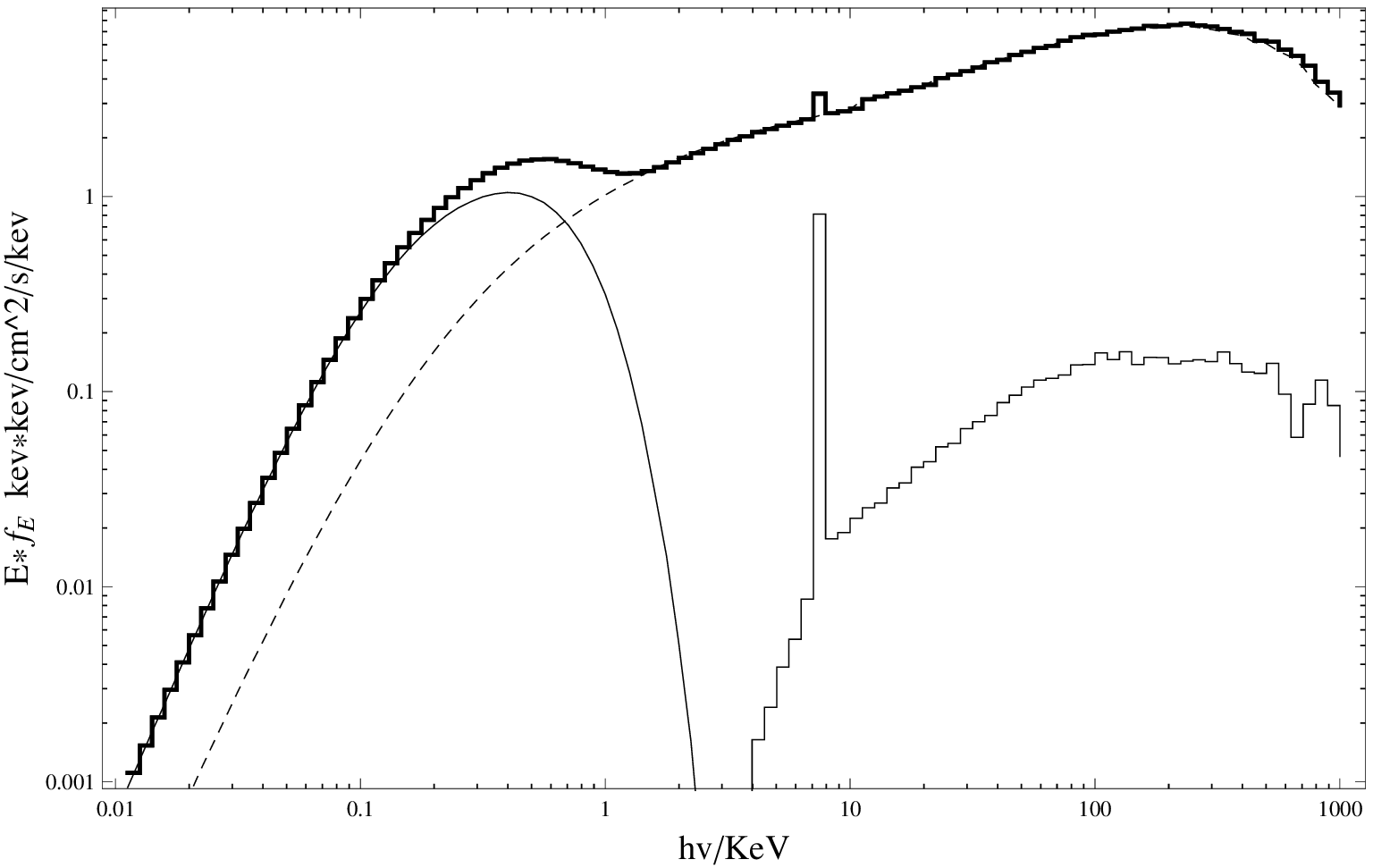}

\end{center}
\caption{Emerged spectrum from the disc-corona system in the cases
of small accretion rate and small black-hole spin (Sample 2). The
total emissive spectrum and its thermal, comptonized and reflective
components are plotted in thick-zigzag, solid, dashed and
thin-zigzag lines, respectively.}
\end{figure*}

\begin{figure*}
\label{fig1} \vspace{0.5cm}
\begin{center}

\includegraphics[height=10cm]{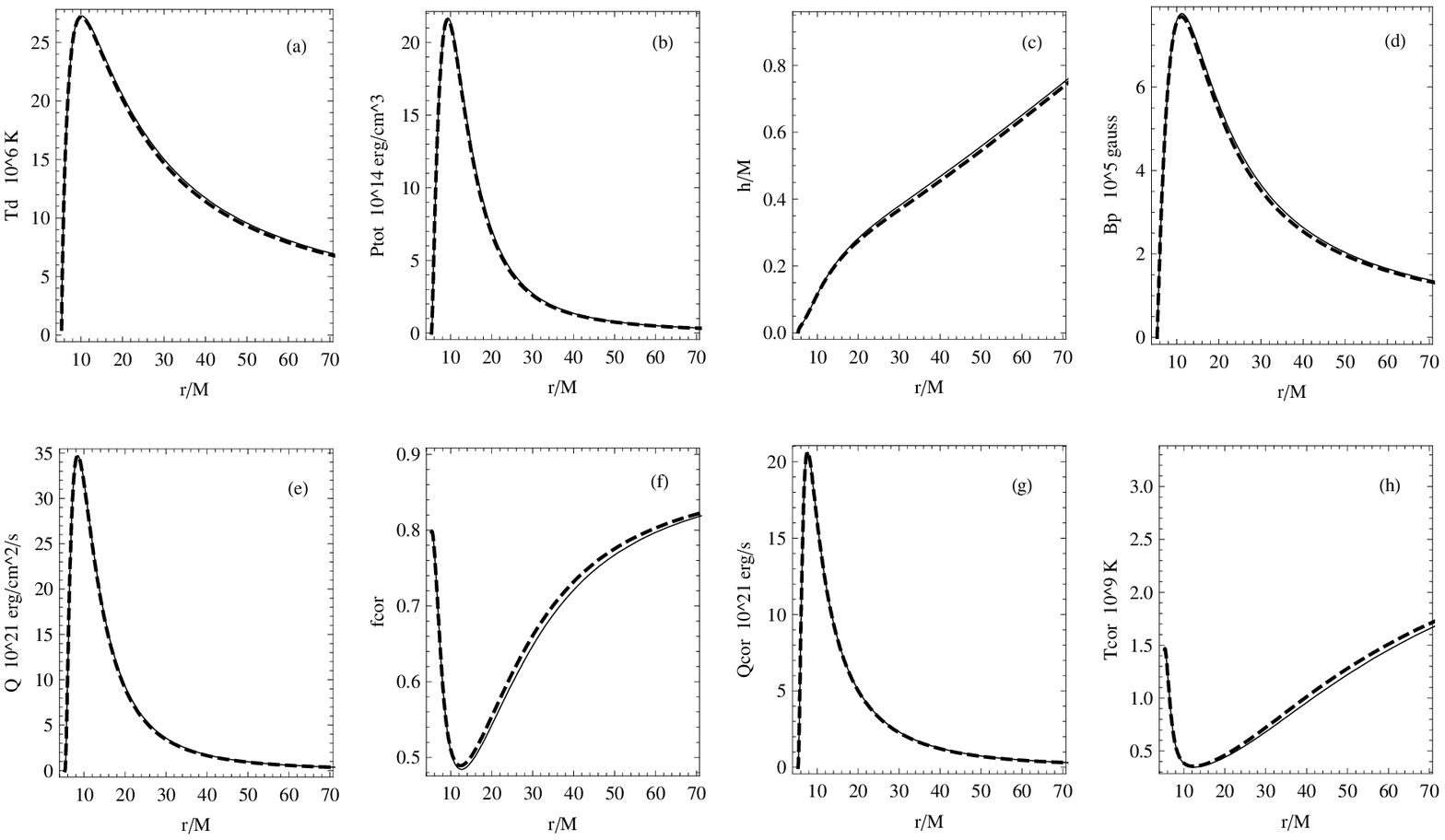}

\end{center}
\caption{A global solution of the magnetic disc-corona system in the
cases of large accretion rate and small black-hole spin (Sample 3).
Quantities are plotted in dashed and solid lines for disc + corona
and disc + corona + MC, respectively. (The dashed and solid lines
almost coincide completely.)}
\end{figure*}

\begin{figure*}
\label{fig1} \vspace{0.5cm}
\begin{center}

\includegraphics[height=8cm]{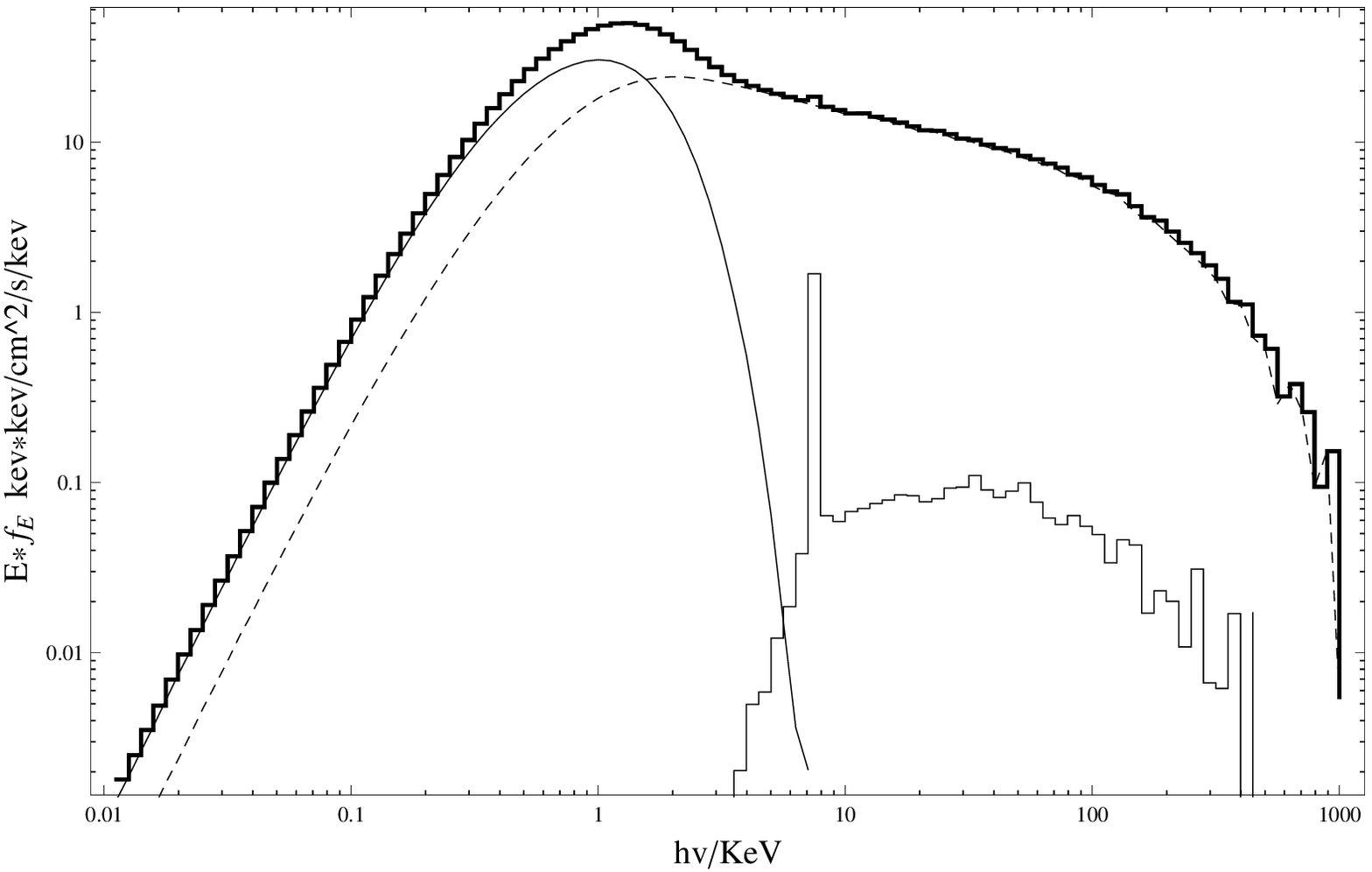}

\end{center}
\caption{ Emerged spectrum from the disc-corona system in the cases
of large accretion rate and small black-hole spin (Sample 3). The
total emissive spectrum and its thermal, comptonized  and reflective
components are plotted in thick-zigzag, solid, dashed and
thin-zigzag lines, respectively.}
\end{figure*}

\begin{figure*}
\label{fig1} \vspace{0.5cm}
\begin{center}

\includegraphics[height=10cm]{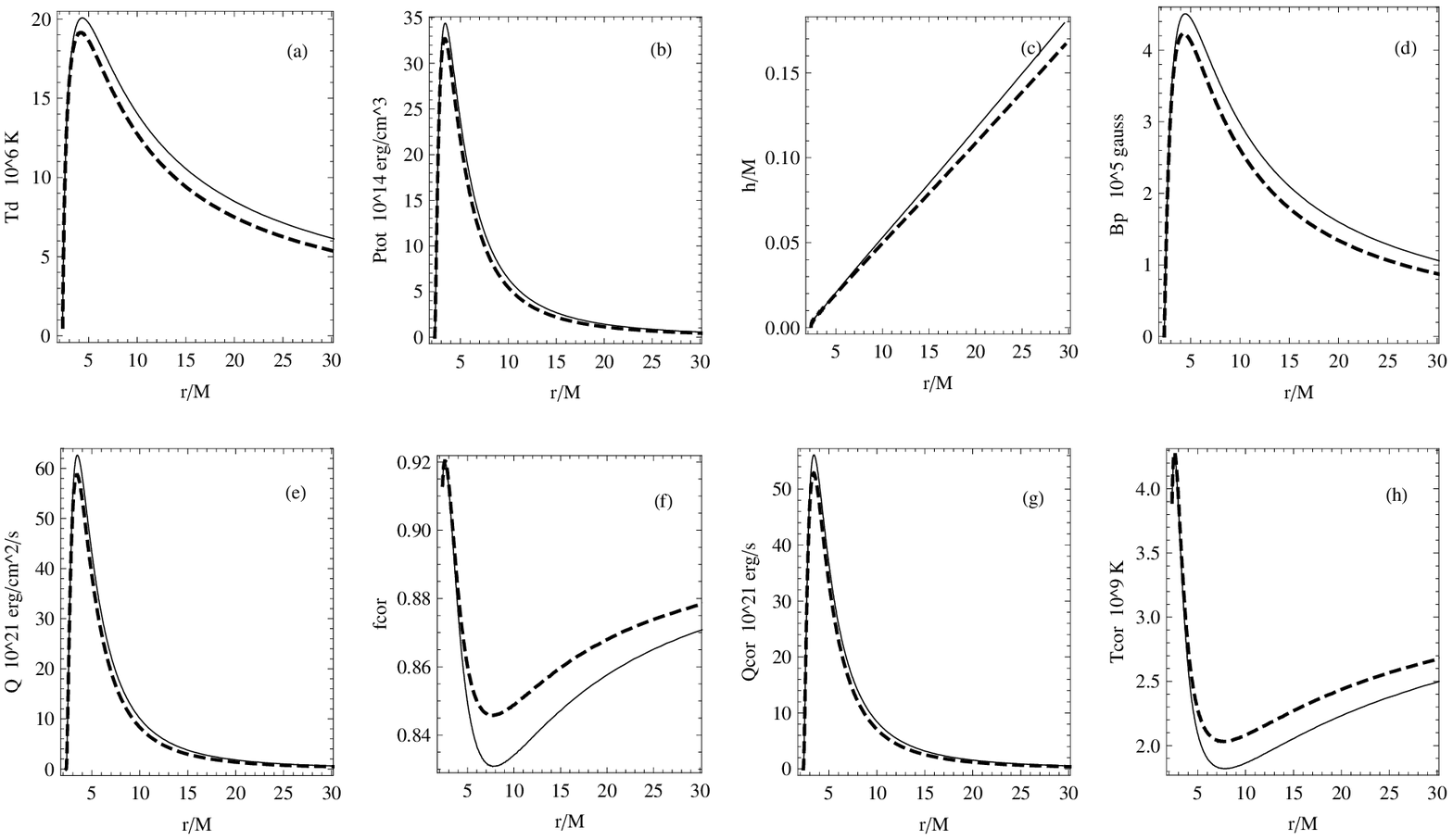}

\end{center}
\caption{A global solution of the magnetic disc-corona system in the
cases of small accretion rate and large black-hole spin (Sample 4).
Quantities are plotted in dashed and solid lines for disc + corona
and disc + corona + MC, respectively. }
\end{figure*}

\begin{figure*}
\label{fig1} \vspace{0.5cm}
\begin{center}

\includegraphics[height=8cm]{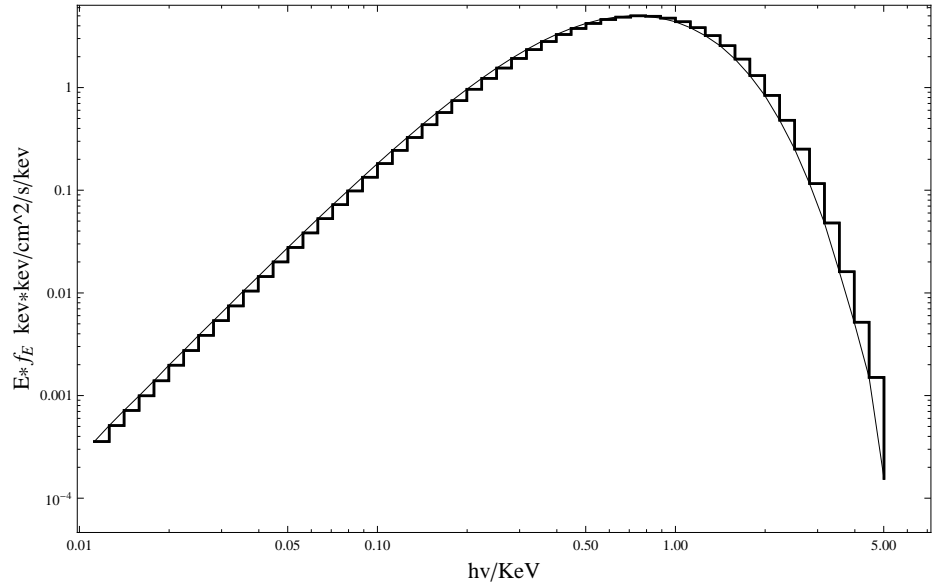}

\end{center}
\caption{Emerged spectrum from the disc-corona system in the cases
of small accretion rate and large black-hole spin (Sample 4). The
total emissive spectrum and its thermal, comptonized and reflective
components are plotted in thick-zigzag, solid, dashed and
thin-zigzag lines, respectively. (Note: The comptonized and
reflective components are too weak and absent from the figure)}
\end{figure*}

\begin{figure*}
\label{fig1} \vspace{0.5cm}
\begin{center}

\includegraphics[height=9cm]{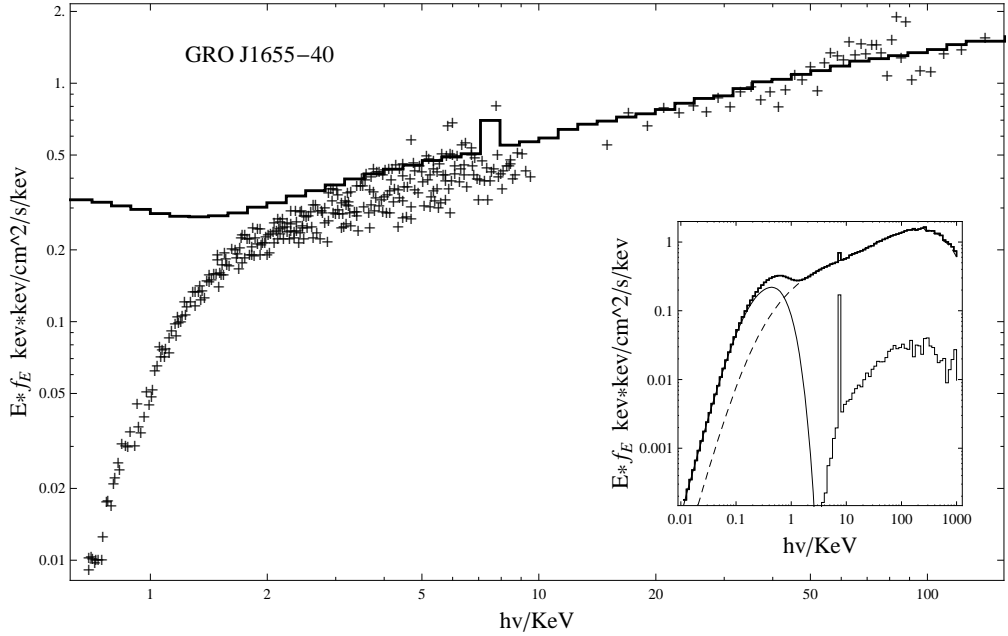}

\end{center}
\caption{GRO J1655-40 in low/hard state (2005.03.06). The Concerned
parameters are listed in Table 1. The embedded small figure is the
emerged spectrum of our model. It is shown that the hard X-ray
spectrum ( $>\sim $ 2 KeV) can be well fitted. It is expected that
the excess in soft X-ray band could be absorbed by the surrounding
medium (cf. Brocksopp et al. 2006).}
\end{figure*}

\begin{figure*}
\label{fig1} \vspace{0.5cm}
\begin{center}

\includegraphics[height=9cm]{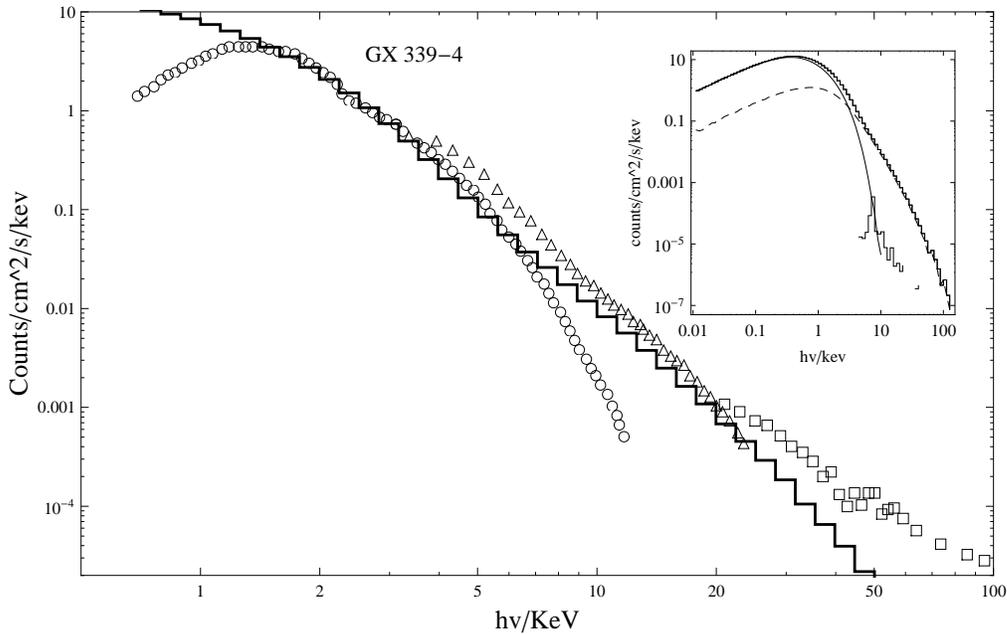}

\end{center}
\caption{GX 339-4 in very high state (2002.09.29). The Concerned
parameters are listed in Table 1. The embedded small figure is the
emerged spectrum of our model. It is shown that the hard X-ray
spectrum ( $>\sim$ 2 KeV) can be well fitted. It is expected that
the excess in soft X-ray band could be absorbed by the surrounding
medium (cf. Miller et al. 2004).}
\end{figure*}

\begin{figure*}
\label{fig1} \vspace{0.5cm}
\begin{center}

\includegraphics[height=9cm]{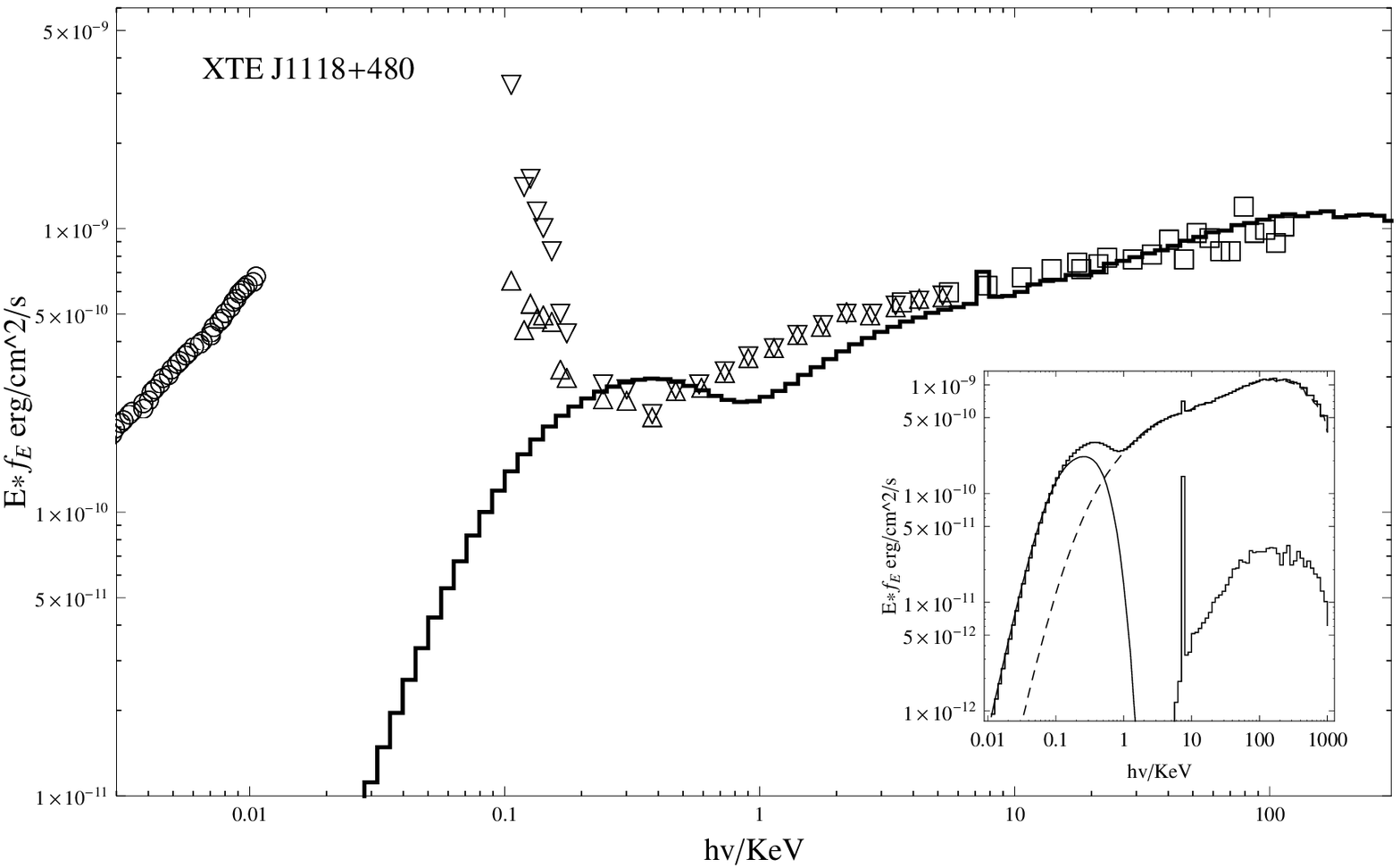}

\end{center}
\caption{XTE J1118+480 in low/hard state. The Concerned parameters
are listed in Table 1. The embedded small figure is the emerged
spectrum of our model. It is shown that the X-ray spectrum ( $>\sim$
0.1 KeV) can be fitted in the rough. The excess in lower energy band
below 0.1 KeV might arise from bolometric radiation from the outer
disc region and synchrotron radiation of jet, which is beyond the
scope of this paper. (cf. Yuan et al. 2005)}
\end{figure*}


\begin{thebibliography}{99}

\bibitem[01]{b01}{{Afshordi N., Paczynski B., 2003, ApJ, 592, 354}}

\bibitem[02]{b02}{{Agol E., Krolik J. H., 2000, ApJ, 528, 161}}

\bibitem[1]{b1}{Balbus S. A., Hawley J. F., 1991, ApJ, 376, 214 }

\bibitem[2]{b2}{Bardeen J. M., Press W. H. \& Teukolsky S. A., 1972, ApJ, 178, 347 }

\bibitem[3]{b3}{Blandford R. D., Znajek R. L., 1977, MNRAS, 179, 433 }

\bibitem[4]{b4}{Blandford R. D., Payne D. G., 1982, MNRAS, 199, 883 (BP82) }

\bibitem[5]{b5}{Brocksopp C., McGowan K. E., Krimm H., Godet O., Roming P.,
Mason K. O., Gehrels N., Still M., Page K., Moretti A., Shrader C.
R., Campana S. \& Kennea J. 2006, MNRAS, 365, 1203 }

\bibitem[6]{b6}{Esin A. A., McClintock J. E. \& Narayan R., 1997, ApJ, 489, 865 }

\bibitem[7]{b7}{Fragile P. C. \& Meier D. L., 2008, arXiv: 0810.1082 {(FM08)}}

\bibitem[71]{b71}{{Gammie C. F., 1999, ApJ, 522, 57} }

\bibitem[72]{b72}{{Garofalo D., Reynolds C. S., 2005, ApJ, 624, 94}}

\bibitem[8]{b8}{Gierlinski M., Zdziarski A. A., Done C., Johnson W. N.,
Ebisawa K., Ueda Y., Phlips F., 1997, MNRAS, 288, 958. }

\bibitem[81]{b81}{{Hirose S., Krolik J. H., De Villiers J.-P., Hawley J. F., 2004, ApJ,
606, 1083 (H04)}}

\bibitem[82]{b82}{{Krolik J. H., 1999, ApJL, 515, 73}}

\bibitem[9]{b9}{Li L.-X., 2002, ApJ, 567, 463 }

\bibitem[10]{b10}{Liu B.-F., Mineshige S. \& Shibata K., 2002, ApJ, 572, L173 (LMS02) }

\bibitem[11]{b11}{Livio M., Ogilvie G. I., Pringle J. E., 1999, ApJ, 512, 100 }


\bibitem[12]{b12}{McClintock, J E, \& Remillard R A 2006. In Compact Stellar X-ray
Sources, ed. WHG Lewin, M, van der Klis, pp. 157¨C214. Cambridge:
Cambridge University Press. (MR06) }

\bibitem[13]{b13}{Meier D. L., New Astron. Rev. 2003, 47, 667 }

\bibitem[14]{b14}{Merloni A., \& Fabian A. C., 2001, MNRAS, 321, 549 (MF02) }

\bibitem[15]{b15}{Miller J. M., Wijnands R., Homan J., Belloni T., Pooley D.,
Corbel S., Kouveliotou C., van der Klis M., Lewin W. H. G., 2001,
ApJ, 563,928 }

\bibitem[16]{b16}{Miller J. M., Fabian A. C., Reynolds C. S., Nowak M. A., Homan J.,
Freyberg M. J., Ehle M., Belloni T., Wijnands R., van der Klis M.,
Charles P. A., Lewin W. H. G., 2004, ApJ, 606, L131 }

\bibitem[17]{b17}{Miller J. M., Raymond J., Reynolds C. S., Fabian A. C.,
Kallman T. R., Homan J., 2008, ApJ, 680, 1359 }

\bibitem[18]{b18}{Moderski R., Sikora M., Lasota J. P., 1997, in ¡°Relativistic Jets
in AGNs¡± eds.M. Ostrowski, M. Sikora, G. Madejski \& M. Belgelman,
Krakow, p.110 }

\bibitem[19]{b19}{Narayan, R., Yi, I., 1994, ApJ, 428, L13 }

\bibitem[191]{b192}{{Paczynski B., 2000, astro-ph/0004129}}

\bibitem[20]{b20}{Penrose R., 1969, 1: 252 }

\bibitem[21]{b21}{Pringle, J.E., 1981, ARA\&A, 19, 137 }

\bibitem[211]{b211}{{Shafee R., Narayan R., McClintock, J. E., 2008, ApJ, 676, 549}}

\bibitem[22]{b22}{Shakura, N. I. \& Sunyaev, R.A. 1973, Astron.Astrophys, 24, 337 }

\bibitem[23]{b23}{Stella L. \& Nosner R., 1984, ApJ, 277, 312 }

\bibitem[24]{b24}{Taam, R., \& Lin, D. N. C. 1984, ApJ, 287, 761 }

\bibitem[241]{b241}{{Tomimatsu A., Takahashi M., 2001, ApJ, 552, 710}}

\bibitem[25]{b25}{Uzdensky D. A., 2004, ApJ, 603, 652 }

\bibitem[26]{b26}{Uzdensky D. A., 2005, ApJ, 620, 889 }

\bibitem[27]{b27}{Wandel, A., \& Liang, E. P. T. 1991, ApJ, 380, 84 }

\bibitem[28]{b28}{Wang D.-X., Xiao K., Lei W.-H., 2002, MNRAS, 335, 655 }

\bibitem[29]{b29}{Wang D.-X., Ma R.-Y., Lei W.-H., Yao G.-Z., 2003, ApJ, 595,
109}

\bibitem[30]{b30}{Wang J.-M., Watarai K. Y. \& Mineshige S., 2004, ApJ, 607, L107 }

\bibitem[31]{b31}{Wilms J., Reynolds C. S., Begelman M. C., Reeves J.,
Molendi, S., Staubert R., Kendziorra E., 2001, MNRAS, 328, L27 }

\bibitem[32]{b32}{Yuan F., Cui W. \& Narayan R., 2005, ApJ, 620, 905 }

\bibitem[33]{b33}{Zdziarski, A. A. 1999, ASP Conf. Ser. 161: High Energy
Processes in Accreting Black Holes, 16 }

\end{thebibliography}
\end{document}